\documentclass[conference]{IEEEtran}
\IEEEoverridecommandlockouts

\usepackage[utf8]{inputenc}
\usepackage[binary-units=true,mode=text]{siunitx}
\usepackage{layouts}
\usepackage[normalem]{ulem}
\usepackage{xspace}
\usepackage{cite}
\usepackage{multirow}
\usepackage{setspace}
\usepackage{tikz}
\usepackage[keeplastbox]{flushend}

\usepackage{xcolor}
\usepackage{makecell}
\usepackage[hyphens]{url}
\usepackage{graphicx}
\usepackage{wasysym}

\usepackage[inline]{enumitem}
\setlist[itemize]{label=\textbullet,topsep=0pt,leftmargin=1em}
\setlist[enumerate]{label=(\alph*)}

\newcommand{\ie}{i.e.,\xspace}
\newcommand{\eg}{e.g.,\xspace}

\newcommand{\cf}{cf.\xspace}
\newcommand{\etal}{et al.\xspace}

\newcommand{\code}[1]{\texttt{\small#1}\normalsize}

\newcommand{\tabgod}{\CIRCLE}
\newcommand{\tabmed}{\LEFTcircle}
\newcommand{\tabbad}{\Circle}

\DeclareSIUnit{\million}{\text{m}}

\usepackage[hidelinks]{hyperref}
\newcommand\copyrighttext{%
	\footnotesize
	\textcopyright~IFIP, 2020.
    This is the author's version of the work. It is posted here by permission of IFIP for your personal use.
	Not for redistribution.
	The definitive version was published in \emph{IFIP NETWORKING 2020}.
}
\newcommand\copyrightnotice{%
\begin{tikzpicture}[remember picture,overlay]
\node[draw,anchor=south,yshift=30pt] at (current page.south) {\parbox{\dimexpr\textwidth-\fboxsep-\fboxrule\relax}{\copyrighttext}};
\end{tikzpicture}%
}

\newif\ifpandoc
\pandocfalse

\def\assignvar#1=#2\relax{\expandafter\gdef\csname btcstat#1\endcsname{#2\xspace}}

\assignvar AsOf=Apr~13, 2020\relax
\assignvar DateFrom=Jan~1, 2009\relax
\assignvar DateFromRecent=Dec~31, 2016\relax
\assignvar Chainsize=\SI[round-mode=places,round-precision=0,per-mode=symbol]{252.80221823230386}{\gibi\byte}\relax
\assignvar Growthrate=\SI[round-mode=places,round-precision=0,per-mode=symbol]{139.03466456402862}{\mebi\byte\per\day}\relax
\assignvar EvalSynctimeMaxVanilla=\SI[round-mode=places,round-precision=0,per-mode=symbol]{5.176890833333333}{\hour}\relax
\assignvar EvalSynctimeMaxVanillaAbstract=\SI[round-mode=places,round-precision=0,per-mode=symbol]{5.176890833333333}{\hour\ifpandoc h\fi}\relax
\assignvar EvalSynctimeMaxCompaction=\SI[round-mode=places,round-precision=0,per-mode=symbol]{45.723375000000004}{\minute}\relax
\assignvar EvalSynctimeMaxCompactionAbstract=\SI[round-mode=places,round-precision=0,per-mode=symbol]{45.723375000000004}{\minute\ifpandoc min\fi}\relax
\assignvar EvalTrafficRecvMaxVanilla=\SI[round-mode=places,round-precision=0,per-mode=symbol]{229.58562617748976}{\gibi\byte}\relax
\assignvar EvalTrafficRecvMaxVanillaAbstract=\SI[round-mode=places,round-precision=0,per-mode=symbol]{229.58562617748976}{\gibi\byte\ifpandoc GiB\fi}\relax
\assignvar EvalTrafficRecvMaxCompaction=\SI[round-mode=places,round-precision=0,per-mode=symbol]{5.0259119185619054}{\gibi\byte}\relax
\assignvar EvalTrafficRecvMaxCompactionAbstract=\SI[round-mode=places,round-precision=0,per-mode=symbol]{5.0259119185619054}{\gibi\byte\ifpandoc GiB\fi}\relax
\assignvar EvalTrafficBothMaxVanilla=\SI[round-mode=places,round-precision=2,per-mode=symbol]{229.62206963924692}{\gibi\byte}\relax
\assignvar EvalTrafficBothMaxVanillaHighlevel=\SI[round-mode=places,round-precision=0,per-mode=symbol]{229.62206963924692}{\gibi\byte}\relax
\assignvar EvalTrafficBothMaxCompaction=\SI[round-mode=places,round-precision=2,per-mode=symbol]{5.028369883354753}{\gibi\byte}\relax
\assignvar EvalTrafficMeanReduction=\SI[round-mode=places,round-precision=2,per-mode=symbol]{92.5045840889717}{\percent}\relax
\assignvar EvalTrafficMeanReductionHighlevel=\SI[round-mode=places,round-precision=0,per-mode=symbol]{92.5045840889717}{\percent}\relax
\assignvar EvalSynctimeMeanReduction=\SI[round-mode=places,round-precision=2,per-mode=symbol]{75.5752880038477}{\percent}\relax
\assignvar EvalSynctimeMeanReductionHighlevel=\SI[round-mode=places,round-precision=0,per-mode=symbol]{75.5752880038477}{\percent}\relax

\assignvar EvalStorageMaxVanilla=\SI[round-mode=places,round-precision=2,per-mode=symbol]{264.2781256828457}{\gibi\byte}\relax
\assignvar EvalStorageMaxVanillaHighlevel=\SI[round-mode=places,round-precision=0,per-mode=symbol]{264.2781256828457}{\gibi\byte}\relax
\assignvar EvalStorageMaxCompaction=\SI[round-mode=places,round-precision=2,per-mode=symbol]{8.854102448560297}{\gibi\byte}\relax
\assignvar EvalStorageMaxCompactionHighlevel=\SI[round-mode=places,round-precision=0,per-mode=symbol]{8.854102448560297}{\gibi\byte}\relax
\assignvar EvalStorageMaxSave=\SI[round-mode=places,round-precision=2,per-mode=symbol]{255.42402323428541}{\gibi\byte}\relax
\assignvar EvalStorageMaxSaveHighlevel=\SI[round-mode=places,round-precision=0,per-mode=symbol]{255.42402323428541}{\gibi\byte}\relax
\assignvar EvalStorageMeanReduction=\SI[round-mode=places,round-precision=2,per-mode=symbol]{85.59882028916773}{\percent}\relax
\assignvar EvalStorageMeanReductionHighlevel=\SI[round-mode=places,round-precision=0,per-mode=symbol]{85.59882028916773}{\percent}\relax
\assignvar EvalStorageHeaderchainSize=\SI[round-mode=places,round-precision=2,per-mode=symbol]{71.52586460113525}{\mebi\byte}\relax
\assignvar EvalStorageHeaderchainPerBlock=\SI[round-mode=places,round-precision=2,per-mode=symbol]{125.00050833333333}{\byte}\relax
\assignvar EvalStorageMeanChaintailSize=\SI[round-mode=places,round-precision=2,per-mode=symbol]{460.8580966723167}{\mebi\byte}\relax
\assignvar EvalStorageMeanChaintailSizeLater=\SI[round-mode=places,round-precision=2,per-mode=symbol]{0.9862961440347136}{\gibi\byte}\relax
\assignvar EvalStorageOpreturnMaxSize=\SI[round-mode=places,round-precision=0,per-mode=symbol]{5384787}{\count}\relax

\newcommand{\name}{CoinPrune\xspace}

\begin{document}

\bstctlcite{IEEEexample:BSTcontrol}

\title{How to Securely Prune Bitcoin's Blockchain}

\author{\IEEEauthorblockN{Roman Matzutt\IEEEauthorrefmark{1}, Benedikt Kalde\IEEEauthorrefmark{1}, Jan Pennekamp\IEEEauthorrefmark{1}, Arthur Drichel\IEEEauthorrefmark{2}, Martin Henze\IEEEauthorrefmark{3}, Klaus Wehrle\IEEEauthorrefmark{1}}
\IEEEauthorblockA{\IEEEauthorrefmark{1}Communication and Distributed Systems, RWTH Aachen University, Germany $\cdot$ \{lastname\}@comsys.rwth-aachen.de}
\IEEEauthorblockA{\IEEEauthorrefmark{2}IT Security Research Group, RWTH Aachen University, Germany $\cdot$ drichel@itsec.rwth-aachen.de}
\IEEEauthorblockA{\IEEEauthorrefmark{3}Cyber Analysis \& Defense, Fraunhofer FKIE, Wachtberg, Germany $\cdot$ martin.henze@fkie.fraunhofer.de}}

\maketitle
\copyrightnotice

\vspace{-1em}
\begin{abstract}

Bitcoin was the first successful decentralized cryptocurrency and remains the most popular of its kind to this day.
Despite the benefits of its blockchain, Bitcoin still faces serious scalability issues, most importantly its ever-increasing blockchain size.
While alternative designs introduced schemes to periodically create snapshots and thereafter prune older blocks, already-deployed systems such as Bitcoin are often considered incapable of adopting corresponding approaches.
In this work, we revise this popular belief and present \emph{\name}, a snapshot-based pruning scheme that is fully compatible with Bitcoin.
\name can be deployed through an opt-in velvet fork, \ie without impeding the established Bitcoin network.
By requiring miners to publicly \emph{announce} and jointly \emph{reaffirm} recent snapshots on the blockchain, \name establishes trust into the snapshots' correctness even in the presence of powerful adversaries.
Our evaluation shows that \name reduces the storage requirements of Bitcoin already by two orders of magnitude today, with further relative savings as the blockchain grows.
In our experiments, nodes only have to fetch and process \btcstatEvalTrafficRecvMaxCompactionAbstract instead of \btcstatEvalTrafficRecvMaxVanillaAbstract of data when joining the network, reducing the synchronization time on powerful devices from currently \btcstatEvalSynctimeMaxVanillaAbstract to \btcstatEvalSynctimeMaxCompactionAbstract, with even more savings for less powerful devices.
\end{abstract}


\IEEEpeerreviewmaketitle

\vspace{-.8em}
\section{Introduction}
\label{sec:introduction}
\vspace{-.1em}

Bitcoin~\cite{2008_nakamoto_bitcoin} and its (public) blockchain, an immutable append-only ledger of financial transactions, constitute an ongoing success story.
Via the blockchain, all nodes can independently verify Bitcoin's history and thus mutually distrusting peers can establish consensus about the correctness of those transactions, which now also fuels applications such as audit systems~\cite{proof_of_existence,namecoin,2017_henze_dcam}, transparency overlays~\cite{2016_chase_transparency_overlays,2017_nikitin_chainiac}, anonymity bootstrapping services~\cite{2020_matzutt_anonboot}, and smart contracts~\cite{2016_wood_ethereum}.

However, Bitcoin also serves as a prime example of the scalability challenges of widely used blockchain systems.
Among these challenges are, \eg limited transaction throughput, high payment verification delays~\cite{2016_croman_blockchain_scalability}, and, most importantly, ever-growing blockchain sizes.
For instance, Bitcoin's blockchain has a size of \btcstatChainsize with a recent average growth rate of \btcstatGrowthrate as of \btcstatAsOf~\cite{2011_blockchain_com_charts}.
These high demands cause individual nodes to \emph{prune} historical blockchain data~\cite{2015_bitcoin_v0_11_pruning}, \ie older payment flows that have been superseded by newer ones.
While such decisions are rational from the user's perspective, they harm Bitcoin's overall network health:
To root trust into Bitcoin's current state, new nodes need to obtain and reverify \emph{all} blockchain data, including all historical data.
Given the decentralization of public blockchains, which are designed to avoid trusted entities, new nodes require independent sources to obtain all bootstrapping information.

While joining nodes would benefit most if every single node held a full copy of the blockchain, already synchronized nodes need to save disk space instead.
To resolve this inherent conflict of interests, alternative designs~\cite{2014_bruce_mini_blockchain,2016_poelstra_mimblewimble,2016_chepurnoy_rollerchain,2019_schoenfeld_pascal} proposed \emph{snapshot-based synchronization}, where new nodes do not verify all historical data but rely on a recent snapshot of the blockchain's state.
However, well-established blockchain systems, such as Bitcoin, have especially high demands to implement performance improvements but are hard to adapt at the same time:
Major changes have to be adopted by a majority of nodes to prevent permanent network partitioning, which has proven difficult in the past~\cite{2017_morgan_scaling_debate}.
Based on these observations, we identify and raise two main questions:
\emph{
\begin{enumerate*}[label=\bfseries(\alph*)]
    \item{how to extend existing blockchain systems with pruning capabilities and}
    \item{how to do so in a secure and trustworthy manner?}
\end{enumerate*}
}

To answer both questions, we first survey approaches to reducing blockchain sizes, arguing that they either are inefficient, insecure, or not deployable to existing systems.
We thus propose \emph{\name}\footnote{Research prototype available at \url{https://github.com/COMSYS/coinprune}}, a block-pruning scheme that is fully compatible with Bitcoin and can be adopted immediately by any subset of nodes.
\name enables joining nodes to synchronize via recent and \emph{trustworthy} snapshots of Bitcoin's state.
By establishing additional trust into these snapshots in a distributed manner, \name drastically unburdens all Bitcoin nodes:
First, new nodes only need to obtain a small fraction of the blockchain.
Further, \name enables all nodes to prune obsolete information \emph{without} affecting the overall network health.
\name establishes trust by periodically having miners cryptographically tie snapshots to the blockchain while other miners \emph{independently reaffirm} the snapshots' correctness.
Thus, assuming an appropriate partial adoption of our scheme, joining nodes may rely on the honest network majority to verify snapshots.
Reaffirmations are ignored by legacy nodes and do not affect block acceptance.
Hence, \name can be deployed mid-operation via a velvet fork~\cite{2017_kiayias_velvet_forks,2018_zamyatin_velvet_forks}.
Consequently, new nodes only require a recent reaffirmed snapshot and subsequent full blocks to synchronize.

Our evaluation shows that full nodes and miners supporting \name can reduce disk space utilization by \btcstatEvalStorageMeanReductionHighlevel.
Still, they are able to help joining nodes synchronize.
In fact, \name drastically improves synchronization performance:
While network traffic generated by joining nodes is reduced by \btcstatEvalTrafficMeanReductionHighlevel, synchronization time drops from \btcstatEvalSynctimeMaxVanilla to \btcstatEvalSynctimeMaxCompaction on powerful devices and even more for less powerful devices.

\section{Bitcoin Overview}
\label{sec:background}

We start off with a primer on Bitcoin before describing its transaction management, blockchain layout with implications on consensus updates, and its bootstrapping process.

\textbf{Bitcoin Primer.}
Bitcoin's~\cite{2008_nakamoto_bitcoin} main contribution was its blockchain, a public and immutable append-only ledger of financial transactions to prevent the double-spending of coins within an untrusted peer-to-peer (P2P) network.
Bitcoin establishes this ledger by bundling pending transactions in cryptographically interlinked, hard-to-create blocks.
The blockchain is jointly maintained by a P2P network of \emph{full nodes} that locally verify pending transactions and blocks, discarding any incorrect information.
Special nodes, the \emph{miners}, invest their computational power to create new blocks by solving a proof-of-work (PoW) puzzle in exchange for freshly minted bitcoins as a reward.
Modifying blocks at a later point becomes increasingly hard as it requires recomputing all subsequent blocks to keep their chaining intact.

\textbf{Transaction Management.}
Each transaction transfers previously received or minted bitcoins to one or more receivers via individual \emph{transaction outputs}.
To prevent double-spending, full nodes have to verify claimed coin ownership for all pending transactions.
For efficiency reasons, all nodes keep track of the current \emph{set of unspent transaction outputs (UTXO set)}.
Thereby, nodes can discard pending transactions that attempt to spend non-existing or already spent bitcoins.
Notably, by default, full nodes prune all spent transactions from their transaction index, \ie if they do not contribute to the UTXO set anymore.
Nevertheless, these nodes retain a full copy of historical blocks to help bootstrap new nodes.

\textbf{Blockchain Layout and Consensus Updates.}
Each Bitcoin block consists of a header and a set of transactions attached to the block via a Merkle tree.
The \SI{80}{\byte}-long header consists of a  version field, the hash value of the block's predecessor for chaining the blocks, the Merkle tree root to cryptographically tie the transactions to the block header, an approximate timestamp, as well as the miner's PoW.
Given the distributed nature of Bitcoin, blockchain \emph{forks}, \ie situations in which the blockchain diverges into more than one potential path forward, can occur either accidentally or with intent.
Accidental forks occur when multiple miners find valid blocks concurrently.
These forks are resolved naturally since one branch is highly likely to grow faster (\ie accumulate more PoW), causing all nodes to abandon other branches in favor of the fastest-growing branch.
Intentional forks are used to \emph{update} existing consensus rules and are traditionally categorized as either \emph{hard forks} or \emph{soft forks}~\cite{2018_zamyatin_velvet_forks}.
While hard forks introduce protocol-breaking changes to the consensus rules, \eg altered block structures, soft forks aim to remain backward-compatible with clients following older consensus rules~\cite{2018_zamyatin_velvet_forks}.
Both paradigms can incur \emph{permanent} blockchain forks depending on whether the majority of nodes accepts or rejects the proposed changes.
Contrarily, multiple works~\cite{2017_kiayias_velvet_forks,2018_zamyatin_velvet_forks} recently investigated \emph{velvet forks}, which aim to allow for the \emph{gradual introduction} of new features without creating permanent forks.
This type of fork augments upgraded blocks in a way that is still valid to legacy nodes, while updated nodes process them in accordance with the changed protocol.

\textbf{Initial Synchronization.}
When a node first joins the Bitcoin network, it needs to obtain its individual view on Bitcoin's current state of consensus, \ie the UTXO set resulting from the blockchain path containing most PoW.
To keep this process fully decentralized and independent from trusted nodes, each node initially establishes eight outgoing connections to random established nodes, called \emph{neighbors}, and downloads the complete blockchain from them.
Due to the separation of headers and transactions, nodes first fetch the \emph{headerchain}, \ie chain of block headers, and simultaneously request full blocks, \ie the corresponding transactions.
While receiving the data, the joining node verifies its correctness by
\begin{enumerate*}
    \item{verifying the blockchain's cryptographic links back to the hard-coded genesis block,}
    \item{keeping track of the amount of performed PoW to remain on the currently valid blockchain,}
    \item{validating transaction sets tied to each block, and}
    \item{by checking the correctness of transactions and replaying them to obtain an up-to-date UTXO set.}
\end{enumerate*}
Even though this information is sufficient to process newly mined blocks, nodes keep a full copy of the blockchain by default.

\section{Impact of Growing Blockchain Sizes}
\label{sec:motivation}

By design, blockchains continuously grow in size and thus eventually reach prohibitive sizes.
For instance, the most popular system, Bitcoin, suffers from a blockchain size of currently \btcstatChainsize with a recent average growth rate of \btcstatGrowthrate as of \btcstatAsOf~\cite{2011_blockchain_com_charts}.
This worrying trend severely impacts the scalability of the overall network.
In addition to increasing disk space requirements, which already today exclude devices such as smartphones from running a full Bitcoin node, we observe negative influences of bandwidth requirements, processing costs, and synchronization times of newly joining nodes.

\textbf{Storage Requirements.}
To retain a decentralized consensus network, Bitcoin requires that enough independent nodes persistently maintain a full blockchain copy to help bootstrap joining nodes (\cf Section~\ref{sec:background}).
However, storing hundreds of Gigabytes of historical blockchain data is both irrational for individual node operators and prohibitive on storage-constrained devices.
In consequence, such devices cannot act as full nodes, and users have to accept weakened security guarantees.

\textbf{Bandwidth Requirements.}
During initial synchronization, each joining node must obtain a full blockchain copy.
Current blockchain sizes already require good Internet connectivity for both the joining node and its serving nodes, potentially causing increased initial synchronization times for joining nodes.
Furthermore, such requirements also put an additional burden onto existing nodes as serving new nodes consumes resources that could otherwise be used for other tasks, \eg gossiping pending transactions or newly mined blocks.

\textbf{Processing Costs.}
In addition to downloading the blockchain, joining nodes also need to verify the blockchain's integrity and locally replay every single transaction to build the UTXO set.
This process consumes excessive amounts of computation power for joining nodes.
Especially, the presence of large numbers of obsolete transactions that do not contribute to the UTXO set wastes valuable resources anymore.

\textbf{Synchronization Time.}
The combination of high bandwidth requirements and high processing costs cause prolonged synchronization times.
While benchmarks using powerful clients report about \SI{5}{\hour} in 2018~\cite{2018_lopp_bitcoin_scalability}, literature already highlighted this issue in 2016 when four days were required to synchronize Amazon EC2 nodes~\cite{2016_croman_blockchain_scalability}.
Naturally, this problem aggravates over time as new blocks are added continuously.

In summary, the ever-growing blockchain heavily impedes the scalability of the overall system for both joining and existing nodes.
Existing nodes are even punished for acting altruistically in the network by helping joining nodes synchronize.
These problems are especially severe for popular systems such as Bitcoin.
In the following, we survey to which extent (newly proposed) schemes attempt to tackle these issues.

\begin{table*}[t]
    \centering
    \caption{Qualitative Comparison of Approaches Improving Storage Requirements and Initial Synchronization}
    \setlength{\tabcolsep}{5pt}
    \scriptsize
    \begin{tabular}{p{0.0cm}l|l||c|c|c|c|c|c|c|c|c}
        &
        \textbf{Name} &
        \textbf{Approach} &
        \textbf{\makecell{Reduce \\ Processing}} &
        \textbf{\makecell{Reduce \\ Traffic}} &
        \textbf{\makecell{Reduce \\ Storage}} &
        \textbf{\makecell{Sync. \\ Time}} &
        \textbf{\makecell{Maintain \\ Security}} &
        \textbf{\makecell{Network \\ Health}} &
        \textbf{\makecell{Server \\ Burden}} &
        \textbf{\makecell{Comple- \\ teness}} &
        \textbf{\makecell{Compat- \\ ibility}} \\
        \hline
        \multirow{9}{*}{\hspace{-0.1cm}\rotatebox{90}{Deployed Solutions}} &
        Hot Wallets~\cite{2012_bitcoin_hot_wallet} &
        Trust Delegation &
        \tabbad\ / \tabgod &
        \tabbad\ / \tabgod &
        \tabbad\ / \tabgod &
        \tabbad\ / \tabgod &
        \tabbad &
        \tabbad &
        \tabgod &
        \tabbad &
        \tabgod \\
        &
        Light Nodes~\cite{2018_bitcoin_light_nodes} &
        Trust Delegation &
        \tabbad\ / \tabgod &
        \tabbad\ / \tabgod &
        \tabbad\ / \tabgod &
        \tabbad\ / \tabgod &
        \tabmed &
        \tabbad &
        \tabgod &
        \tabbad &
        \tabgod \\
        &
        ``Ultimate Compression''~\cite{2012_reiner_ultimate_compression} &
        Trust Delegation &
        \tabbad\ / \tabgod &
        \tabbad\ / \tabgod &
        \tabbad\ / \tabgod &
        \tabbad\ / \tabgod &
        \tabbad &
        \tabgod &
        \tabbad &
        \tabgod &
        \tabgod \\
        &
        DB Improvements~\cite{2013_bitcoin_v0_8_0_db_change,2017_bitcoin_v0_15_utxo_changed} &
        Data Management &
        \tabmed &
        \tabbad &
        \tabbad &
        \tabmed &
        \tabgod &
        \tabgod &
        \tabgod &
        \tabgod &
        \tabgod \\
        &
        Index Pruning~\cite{2013_bitcoin_v0_8_0_db_change} &
        Data Management &
        \tabmed &
        \tabbad &
        \tabmed &
        \tabmed &
        \tabgod &
        \tabgod &
        \tabgod &
        \tabmed &
        \tabgod \\
        &
        Headers-first Download~\cite{2015_bitcoin_v0_10_synchronization} &
        Data Management &
        \tabbad &
        \tabbad &
        \tabbad &
        \tabmed &
        \tabgod &
        \tabgod &
        \tabgod &
        \tabgod &
        \tabmed \\
        &
        Assume-valid Blocks~\cite{2010_nakamoto_v0_3_2_checkpoints,2017_bitcoin_v0_14_assumevalid} &
        Skip Verification &
        \tabgod &
        \tabbad &
        \tabbad &
        \tabmed &
        \tabgod &
        \tabgod &
        \tabgod &
        \tabgod &
        \tabgod \\
        &
        Block Pruning~\cite{2015_bitcoin_v0_11_pruning} &
        Simple Block Pruning &
        \tabbad &
        \tabbad &
        \tabgod &
        \tabbad &
        \tabgod &
        \tabbad &
        \tabgod &
        \tabmed &
        \tabgod \\
        &
        Ethereum Fast Sync~\cite{2015_szilagyi_ethereum_fast_sync} &
        State-based Sync. &
        \tabgod &
        \tabbad &
        \tabbad &
        \tabgod &
        \tabgod &
        \tabgod &
        \tabgod &
        \tabgod &
        \tabbad \\
        \hline
        \multirow{7}{*}{\hspace{-0.1cm}\rotatebox{90}{Related Work}} &
        Selective Pruning~\cite{2018_palm_selective_pruning} &
        Simple Block pruning &
        \tabmed  &
        \tabmed  &
        \tabgod  &
        \tabmed  &
        \tabbad  &
        \tabgod  &
        \tabbad  &
        \tabgod  &
        \tabbad \\
        &
        Rollerchain~\cite{2016_chepurnoy_rollerchain} &
        State-based Sync. &
        \tabgod &
        \tabgod &
        \tabgod &
        \tabgod &
        \tabgod &
        \tabgod &
        \tabmed &
        \tabmed &
        \tabbad \\
        &
        Marsalek \etal~\cite{2019_marsalek_compression} &
        State-based Sync. &
        \tabbad\ / \tabgod &
        \tabbad\ / \tabgod &
        \tabbad\ / \tabgod &
        \tabgod &
        \tabgod &
        \tabmed &
        \tabmed &
        \tabmed &
        \tabbad \\
        &
        Mini Blockchain Scheme~\cite{2014_bruce_mini_blockchain} &
        Balance-based Sync. &
        \tabgod &
        \tabgod &
        \tabgod &
        \tabgod &
        \tabgod &
        \tabmed &
        \tabmed &
        \tabbad &
        \tabbad \\
        &
        Mimblewimble~\cite{2016_poelstra_mimblewimble} &
        Balance-based Sync. &
        \tabbad &
        \tabmed &
        \tabmed &
        \tabbad &
        \tabgod &
        \tabgod &
        \tabgod &
        \tabbad &
        \tabbad \\
        &
        Pascal~\cite{2019_schoenfeld_pascal} &
        Balance-based Sync. &
        \tabgod &
        \tabgod &
        \tabgod &
        \tabgod &
        \tabmed &
        \tabbad &
        \tabgod &
        \tabmed &
        \tabbad \\
        &
        Vault~\cite{2019_leung_vault} &
        Balance-based Sync. &
        \tabgod &
        \tabgod &
        \tabgod  &
        \tabgod  &
        \tabgod  &
        \tabgod  &
        \tabgod  &
        \tabmed  &
        \tabbad \\
		\hline
        &
        \name (our approach) &
        State-based Sync. &
        \tabgod &
        \tabgod &
        \tabgod &
        \tabgod &
        \hphantom{*}\tabgod* &
        \hphantom{*}\tabgod* &
        \tabmed &
        \tabmed &
        \tabgod \\
    \end{tabular}

    \vspace{0.6em}
    \hspace{3.4em}{%
    \itshape
    $\Box\, /\,\Box$: Distinction Full Nodes / Light Nodes
    \hfill
    $*$: Dependent on honest majority among adopters (\cf Section~\ref{sec:security})
    }\hspace{2.4em}
    \label{tab:comparison}
    \vspace{-2em}
\end{table*}

\section{The Current State of Blockchain Pruning}
\label{sec:sota}

Blockchains, especially public ones, have long suffered from their limited scalability.
Consequentially, developers have tackled these scalability challenges from different perspectives.
In this section, we survey current state-of-the-art measures deployed to existing systems as well as alternative blockchain designs that focus on reducing storage requirements and improving the bootstrapping.
Other works that consider blockchain data management, but are not covered here explicitly, include analyses of blockchain data~\cite{2013_dorit_transaction_analysis,2013_meiklejohn_anonymity_analysis,2017_bartoletti_opreturn,2018_matzutt_contents,2017_sward_contents} and the UTXO set~\cite{2018_delgado_segura_utxo}, approaches to prevent illicit content from being engraved into the blockchain~\cite{2017_ateniese_redactable_blockchain,2017_puddu_mu_chain,2018_matzutt_thwarting,2019_deuber_redactable_blockchain,2019_florian_erase,2019_dorri_mofbc}, sharding approaches~\cite{2016_luu_elastico,2018_zamani_rapidchain,2018_kokoris_omniledger}, and lightweight payment schemes~\cite{2016_poon_lightning,2017_green_bolt}.

\subsection{Survey Criteria and Methodology}
\label{sec:sota:criteria}

We qualitatively assess the applicability and effectiveness of related approaches based on their
\begin{enumerate*}
    \item{scalability improvements,}
    \item{whether they maintain sufficient levels of security,}
    \item{their impact on the overall network,}
    \item{potential impact on blockchain queryability,}
    \item{and their compatibility with already established public blockchain systems such as Bitcoin.}
\end{enumerate*}

Regarding \emph{scalability improvements}, we separately consider processing, traffic, and storage.
Based on these improvements, we discuss the overall impact on synchronization times for joining nodes.
We resort to a qualitative assessment of the presented approaches, as most works do not present performance benchmarks.
We comment on the approaches' \emph{security} by discussing whether or not these proposals actively weaken security guarantees of the base system.
Furthermore, we discuss the \emph{impact on the overall network} by considering the network health, \ie the dependency on especially altruistic nodes, and the potential overhead the approaches may introduce for already synchronized nodes.
Then, we assess how the proposed schemes may impact the \emph{blockchain queryability}, \eg the capability of querying historical transactions or augmented transactions such as Bitcoin's \code{OP\_RETURN} transactions.
Finally, we survey their \emph{compatibility} with already deployed blockchain systems.
We summarize our results in Table~\ref{tab:comparison}.

\subsection{Measures Deployed in Existing Blockchain Systems}
\label{sec:sota:updates}

The increased popularity of cryptocurrencies forced their developers to tackle rising scalability issues.
In this section, we present measures taken either by users locally or by blockchain developers network-wide.
Our discussion is based on the reference implementations (Bitcoin Core and Ethereum's \code{geth}, respectively) where appropriate.
Overall, we identify approaches based on \emph{trust delegation}, \emph{skipping verification} steps, \emph{improving data management} with the special case of \emph{block pruning}, and \emph{state-based synchronization}.

\textbf{Trust Delegation.}
Users can delegate their trust into the blockchain's correctness to third parties if they cannot operate a full node, \eg when using a constrained device for issuing transactions.
Using \emph{hot wallets}~\cite{2012_bitcoin_hot_wallet}, users essentially outsource all fund management to a trusted third party, enabling the service provider to issue transactions on their behalf.
Similarly, \emph{light nodes}~\cite{2018_bitcoin_light_nodes} outsource blockchain verification to other full nodes, but they manage their wallet locally using simplified payment verification (SPV)~\cite{2008_nakamoto_bitcoin}.
These approaches vastly improve the performance of clients, only put a negligible burden on the full nodes, and they are actively used.
However, they only seize the resources of other nodes and do not contribute positively to the overall network.
Contrarily, trust-delegating nodes heavily rely on a backbone network of full nodes for both trust and relevant information and prohibit local verifiability.
The never deployed Ultimate Compression scheme~\cite{2012_reiner_ultimate_compression} aimed at bootstrapping light nodes with the current UTXO set but requires full nodes to store and transmit a searchable representation of the UTXO set in addition to its full blockchain copy, putting extra burden on the full nodes.
Furthermore, this scheme requires an additional blockchain to establish trust in the transmitted UTXO set.

\textbf{Improving Data Management.}
Increasing blockchain sizes necessitate optimized data management, either for looking up relevant information or for efficiently bootstrapping new nodes.
To this end, Bitcoin Core has historically changed its \emph{underlying database} system~\cite{2013_bitcoin_v0_8_0_db_change} and the internal layout of its UTXO set~\cite{2017_bitcoin_v0_15_utxo_changed}.
Furthermore, full nodes \emph{locally prune} irrelevant entries from their \emph{transaction index}~\cite{2013_bitcoin_v0_8_0_db_change}.
While the raw blockchain data is still persisted, historical information is not queryable anymore.
Network-related optimizations mainly engulf a revised \emph{header-first download} of the blockchain~\cite{2015_bitcoin_v0_10_synchronization}.
Verifying the headerchain is sufficient to ensure the blockchain's integrity.
Since transactions can be decoupled from their block's headers (\cf Section~\ref{sec:background}), nodes can now download and verify full blocks in parallel with only minor and local upgrading incompatibilities~\cite{2015_bitcoin_v0_10_synchronization}.
Nodes can further \emph{limit their block-serving bandwidth}~\cite{2016_bitcoin_v0_12_limit_upload}, and they can relay \emph{compact representations} of newly mined blocks, thereby avoiding transmitting known-but-pending transactions redundantly~\cite{2016_bitcoin_v0_13_segwit}.
However, header-first download still requires to transfer and process all blockchain data during initial synchronization, only its distribution is more efficient.

\textbf{Skipping Verification.}
Bitcoin's reference implementation started early on to avoid reverifying transactions from very old blocks.
Using hard-coded \emph{checkpoint blocks} at first~\cite{2010_nakamoto_v0_3_2_checkpoints}, Bitcoin has recently shifted to use configurable \emph{assumed-valid blocks}~\cite{2017_bitcoin_v0_14_assumevalid}.
The reasoning here is that invalid transactions would have been rejected by the network earlier, and thus older transactions with many confirmations are believed to be correct.
By skipping assumed-valid blocks, joining nodes can avoid the costly signature verification of large portions of the blockchain at negligible security risks.
However, joining nodes still download the complete blockchain to replay all historical transactions to create an up-to-date UTXO set.

\textbf{Simple Block Pruning.}
To counter increasing storage requirements, Bitcoin users have the option to completely \emph{prune raw blockchain data}~\cite{2015_bitcoin_v0_11_pruning} after a \emph{full} initial synchronization.
This step allows nodes to forget all historical blockchain data at the cost of its queryability.
In contrast to local index pruning, block pruning is detrimental to the network health as block-pruning nodes are incapable of bootstrapping new nodes.

\textbf{State-based Synchronization.}
While Bitcoin focuses on financial transactions, other cryptocurrencies, such as Ethereum~\cite{2016_wood_ethereum}, are also capable of executing smart contracts.
Naturally, those cryptocurrencies have more complex state layouts as the full nodes need to keep track of every smart contract's state.
Consequentially, Ethereum uses \emph{Fast Sync}~\cite{2015_szilagyi_ethereum_fast_sync}, which enables joining nodes to download a recent state and thereby avoids replaying all historical information.
However, Ethereum still values the queryability of historical data, and thus joining nodes also download and persist all blocks, but do not have to process them during initial synchronization.
In contrast to Bitcoin's proposal for Ultimate Compression, Fast Sync remains secure since Ethereum, by default, cryptographically ties its current state to each block~\cite{2016_wood_ethereum}.
Thus, nodes can verify the correctness of their obtained state directly via Ethereum's blockchain.
Since other cryptocurrencies lack these header fields, Fast Sync is not immediately portable.

\textbf{Takeaway.}
Developers have tackled the scalability issues of blockchain systems from different perspectives.
However, all approaches have either limited efficiency, questionable security properties, are detrimental to network health, or are not portable to a variety of already deployed systems.

\subsection{Proposed Block-Pruning Schemes}
\label{sec:sota:proposals}

The  insufficiency of post-deployment pruning schemes inspired various \emph{alternative blockchain designs}, promising better scalability than established systems.
We identify alternative designs that refine mere block-pruning schemes as well as designs proposing state-based or balance-based synchronization.

\textbf{Simple Block Pruning.}
Palm \etal~\cite{2018_palm_selective_pruning} present a distributed block-pruning scheme for established nodes in permissioned blockchains, \ie blockchains jointly maintained by a fixed set of mutually known parties.
A dedicated initiator defines a pruning algorithm that must be executed by all nodes to identify and prune now-irrelevant transactions in a way that all relevant data is still retrievable from other nodes.
However, this approach focuses on permissioned blockchains and requires a dedicated initiator.
Hence, the approach is inapplicable to public settings, which are open to unknown or unauthenticated parties, both for security and compatibility reasons.

\textbf{State-based Synchronization.}
Similarly to Ethereum Fast Sync and inheriting its advantages and disadvantages, Rollerchain~\cite{2016_chepurnoy_rollerchain} proposes state-based initial synchronization.
However, Rollerchain values performance over complete queryability, thereby significantly decreasing bootstrapping overhead as old information does not need to be transmitted and stored.
Similarly, Marsalek \etal~\cite{2019_marsalek_compression} propose a state-based synchronization based on Bitcoin but abandon compatibility by rejecting blocks that have invalid states attached.

\textbf{Balance-based Synchronization.}
A special class of state-based block-pruning schemes simplifies the structure of what constitutes a state to allow for more efficient representations and updates~\cite{2014_bruce_mini_blockchain,2016_poelstra_mimblewimble,2019_schoenfeld_pascal,2019_leung_vault}.
Typically, these schemes only keep track of existing accounts and their balances.
The Mini-Blockchain scheme~\cite{2014_bruce_mini_blockchain} replaces Bitcoin's UTXO set with an account tree that is cryptographically tied to each mined block.
Joining nodes obtain the headerchain and a recent account tree to synchronize, before fully processing a tail of full blocks to preserve PoW-based security.
However, such a scheme expects established nodes to compute slices of the recent account tree on demand, without commenting on the availability of all required data to rewind the account tree accordingly within the network.
Mimblewimble~\cite{2016_poelstra_mimblewimble} follows a similar approach, but emphasizes confidential transactions at the cost of synchronization performance as joining nodes have to obtain and verify rangeproofs for unspent funds~\cite{2016_poelstra_mimblewimble}.
Through their balance-based approach, both schemes limit the expressiveness of transactions.
To overcome this limitation, Pascal~\cite{2019_schoenfeld_pascal} defines SafeBoxes as a replacement for mere account trees.
SafeBoxes permit the generation of a limited number of accounts per block and are designed to enable upper-layer applications, but the limited availability of account spots is conceptually detrimental to network health.
Finally, Vault~\cite{2019_leung_vault} builds on Algorand~\cite{2017_gilad_algorand} to enable the distribution of fragments of recent states across the network to reduce the per-node storage requirements.
Therefore, Vault is inapplicable as an aid for existing, simpler cryptocurrencies.

\textbf{Takeaway.}
Alternative blockchain designs have shown that incorporating cryptographic ties to recent state objects are a promising means to establish trust in state-based blockchain synchronization processes.
However, extending existing systems with such capabilities immediately results in hard forks, which are difficult to deploy and thus highly debated~\cite{2017_morgan_scaling_debate}.

\section{Requirements for Secure Block Pruning}
\label{sec:requirements}

Blockchain systems require that sufficiently many nodes maintain a \emph{full local copy} of the blockchain (\cf Section~\ref{sec:background}).
While this initial design becomes massively burdening for these nodes as well as joining nodes, multiple approaches to fully pruning historical data have been proposed (\cf Section~\ref{sec:sota:proposals}).
However, none of these approaches can be adapted to directly provide similar optimizations for established systems (e.g., Bitcoin) without provoking major incompatibilities.
To realize \emph{fully compatible} extensions for the network-wide pruning of obsolete information in existing cryptocurrencies, while \emph{maintaining already established security levels}, we identify the following requirements and design goals:

\newcounter{DesignGoals}
\refstepcounter{DesignGoals}\label{goal:scalability}
\newcommand{\goalscalability}{\textbf{(G\ref{goal:scalability})}\xspace}
\textbf{\goalscalability Scalability.}
To be effective, pruning schemes must provide improvements for \emph{all} metrics discussed in Section~\ref{sec:motivation}, \ie storage and bandwidth demands for joining and block-serving nodes, processing costs, and synchronization time.

\refstepcounter{DesignGoals}\label{goal:correctness}
\newcommand{\goalcorrectness}{\textbf{(G\ref{goal:correctness})}\xspace}
\textbf{\goalcorrectness Correctness.}
Starting from the initial genesis block, each joining node must obtain the same internal state, with or without the block-pruning scheme enabled, to ensure that the network's consensus about accepted transactions is kept intact.
In particular, the node must learn about all accepted, non-obsolete events, and it must not accept any false events.

\refstepcounter{DesignGoals}\label{goal:verifiability}
\newcommand{\goalverifiability}{\textbf{(G\ref{goal:verifiability})}\xspace}
\textbf{\goalverifiability Verifiability.}
As security is a top priority, our pruning scheme must keep joining nodes able to verify the correctness of the synchronization process even in the presence of adversaries.
Here, we require that the block-pruning scheme does not reduce the security of the overall blockchain system. 

\refstepcounter{DesignGoals}\label{goal:compatibility}
\newcommand{\goalcompatibility}{\textbf{(G\ref{goal:compatibility})}\xspace}
\textbf{\goalcompatibility Compatibility.}
Popular and long-living blockchain systems are especially affected by scalability limitations.
Instead of proposing new systems (\cf Section~\ref{sec:sota:proposals}), all changes should be applicable to existing blockchains, especially Bitcoin, even during operation.
Preferably, the scheme is opt-in, \eg as achieved via velvet forks (\cf Section~\ref{sec:background}).

\section{\name Design}
\label{sec:design}

\begin{figure}[t]
    \centering
    \includegraphics{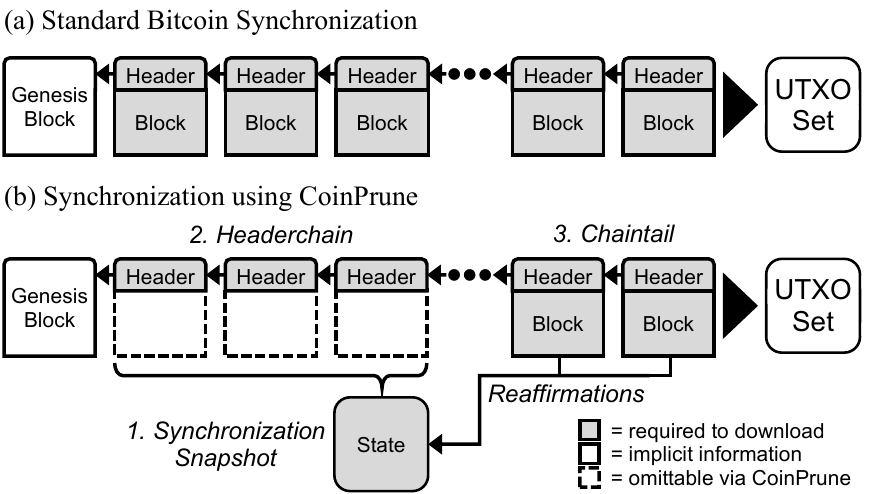}
    \caption{
        High-level design overview of \name.
        Instead of downloading and verifying all blockchain data, joining nodes obtain a recent snapshot in a trustworthy manner due to its on-chain reaffirmations by multiple miners.
    }
    \label{fig:design:overview}
    \vspace{-1.5em}
\end{figure}

To address the challenges resulting from the ever-increasing size of existing blockchains, we present \emph{\name}, our secure, state-based block-pruning scheme that is gradually deployable without protocol-breaking changes, \eg to Bitcoin.
After giving an overview of \name, we first present how nodes using our scheme coordinate via Bitcoin's blockchain and then present how joining nodes can bootstrap.

\subsection{\name Overview}
\label{sec:design:overview}

\name is designed to transfer scalability improvements of novel blockchain designs (\cf Section~\ref{sec:sota:proposals}) to Bitcoin while keeping compatibility a priority.
We now describe how \emph{\name nodes}, \ie Bitcoin nodes additionally supporting \name, jointly maintain recent snapshots on the blockchain, explain how new nodes can bootstrap securely using those snapshots, and outline benefits for the whole network.

\textbf{Snapshot Maintenance.}
\name nodes periodically create \emph{snapshots} of their current UTXO set.
These snapshots are served to joining nodes instead of the entirety of historical blockchain data for reduced storage, bandwidth, and processing requirements~\goalscalability.
They are tied to the current \emph{block height}, \ie the position of the most recent block in the blockchain, and contain a well-ordered UTXO set for synchronization~\goalcorrectness and verification~\goalverifiability purposes, respectively.
To prevent malicious nodes from distributing incorrect snapshots, \eg in an attempt to multiply their unspent funds, \name requires snapshots to be \emph{publicly announced} to the blockchain by referencing a cryptographic \emph{identifier} of each snapshot on-chain.
\name-supporting miners place these announcements in their blocks' \emph{coinbase transactions}, which miners issue to mint new coins.
By utilizing an existing field in coinbase transactions, which may contain \SI{100}{\byte} of arbitrary data, we keep \name Bitcoin-compatible~\goalcompatibility.
Other \name miners independently do the same, which causes nodes deriving snapshots from the same UTXO set to \emph{mutually reaffirm} that snapshot's validity.
This approach creates positive-only feedback, \ie wrong snapshots are not rejected but tolerated and outpaced by valid reaffirmations given an honest majority of \name miners.

\textbf{Bootstrapping Nodes.}
Instead of downloading all historical blockchain data, a joining node can securely bootstrap in three steps, as shown in Figure~\ref{fig:design:overview}:
First, the node obtains a \emph{recent snapshot} either from its neighbors or through a snapshot-offering third party.
If opting for P2P-based snapshot acquisition, the node downloads the snapshot that advocated by most neighbors, given an absolute majority for one snapshot.
Second, the node downloads the \emph{headerchain}, \ie the interconnected and lightweight block headers, to learn about the blockchain branch with the most PoW in it.
Third, the node downloads the \emph{chaintail}, \ie the full blocks starting from the snapshot's block height.
Via the chaintail, the joining node can
\begin{enumerate*}
    \item{catch up with recent transactions, and}
    \item{inspect the full blocks for snapshot reaffirmations.}
\end{enumerate*}
If the joining node observes sufficiently many reaffirmations of the snapshot, it accepts the snapshot and concludes initial synchronization.
Otherwise, the node starts over by discarding the insecure snapshot and reconnecting to a new set of neighbors.

\textbf{Global Block Pruning.}
Since joining nodes can securely bootstrap from the headerchain, the snapshot, and the chaintail, \emph{all} \name nodes may now \emph{safely prune historical blocks} prior to the snapshot.
As new snapshots are reaffirmed periodically, nodes may also prune aging snapshots as well without hurting the network health.
Single \emph{archival nodes} may still keep a full blockchain copy to retain full and reliable queryability of historical data.

\subsection{Adapted Data Management}%
\label{sec:design:data}

To understand \name in detail, we discuss the layout of its snapshots and the required changes to nodes' local data management stemming from the pruning of historical blocks.

\textbf{Snapshot Creation.}
Each snapshot corresponds to a specific block height, meaning that it represents a serialization of the UTXO set obtained from processing all blocks up to and including that height.
A \name snapshot consists of a simple header and multiple chunks of serialized UTXOs, and it is referenced on-chain by a cryptographic identifier.
The header holds the snapshot's corresponding block height, that block's identifier, and the number of chunks in the snapshot.
The identifier is a special hash value created over the snapshot's header and chunks to uniquely represent the snapshot in a succinct manner.
First, the header and each chunk are hashed individually using Bitcoin's HASH256 function (SHA256 applied twice).
Then, the snapshot identifier is the hash value of the concatenation of these hash values.
Using this simple snapshot serialization, joining nodes are immediately aware of all available chunks and can independently request individual chunks from different neighbors in parallel.
Further, we limit chunk sizes to \SI{1}{\mega\byte} akin to Bitcoin's maximum block size.

\textbf{Persisted Information.}
By shifting to snapshot-based synchronization, nodes may now prune historical full blocks.
However, the nodes must remain capable of serving the full headerchain to joining nodes.
Before pruning blocks, these nodes thus need to persist some information currently held by Bitcoin's block index.
These are block identifiers, headers, block heights, the amount of PoW, the number of transactions, and the block's timestamp.
Persisting this data, a recent snapshot, and the chaintail of not-yet-prunable full blocks is now sufficient to securely bootstrap joining nodes.

\begin{figure}[t]
    \centering
    \includegraphics{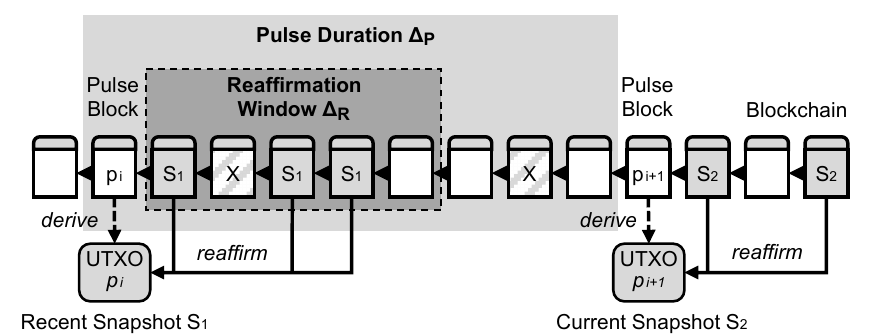}
    \vspace{-1.5em}
    \caption{
        Our pulse-based coordination triggers the creation of new snapshots.
        Any invalid or delayed reaffirmations are ignored.
    }
    \label{fig:design:slotting}
    \vspace{-1.5em}
\end{figure}

\subsection{Coordination of \name Nodes}%
\label{sec:design:coordination}

\name relies on an honest majority of snapshot-serving nodes to mutually reaffirm recent snapshots' correctness to provide a trust anchor for joining nodes.
To this end, these established nodes must agree on when to create snapshots and how to publish reaffirmations.
As shown in Figure~\ref{fig:design:slotting}, \name uses \emph{pulse blocks $p_i$} issued in constant intervals $\Delta_P$, \eg every \num{10000} blocks, and synchronize the snapshot creation and reaffirmation activities among \name nodes.
Through $\Delta_P$, \name can tune the frequency of snapshot creation and thereby adjust processing and storage overheads for \name nodes.
Nodes create a new snapshot whenever a pulse block has been mined.
They base the snapshot on the pulse block's corresponding UTXO set.
Subsequently, all \name miners reaffirm the new snapshot in blocks they mine until the shorter \emph{reaffirmation window $\Delta_R$} expires.
To reaffirm a snapshot, the miners add the snapshot's identifier to their blocks' coinbase fields.
While not all Bitcoin blocks mined during $\Delta_R$ must contain a reaffirmation and individual blocks may reaffirm invalid snapshots, an honest majority among \name miners ensures that the valid, new snapshot will accumulate reaffirmations the fastest.
Nodes \emph{accept} the snapshot with the most reaffirmations if that snapshot was reaffirmed at least $k$ times during $\Delta_R$.
This acceptance threshold $k$ protects nodes from accepting snapshots reaffirmed by individual adversaries with low mining power in times of low overall \name participation.
Consequently, all nodes can safely ignore any reaffirmations that are outpaced by reaffirmations of another snapshot or that lie outside of $\Delta_R$.
If no snapshot reaches $k$ reaffirmations during $\Delta_R$, this pulse is invalid, and pruning is delayed until the next pulse starts.

\subsection{Bootstrapping New Nodes}%
\label{sec:design:bootstrapping}

The reaffirmations periodically published on Bitcoin's blockchain allow joining nodes to bootstrap as follows.
First, the node obtains a recent snapshot.
The node can either acquire a recent snapshot through external means, \eg mirror servers, or ask its neighbors for the most recent snapshots they are aware of using off-chain P2P requests.
Second, the joining node downloads and verifies the headerchain from its neighbors to learn about the blockchain branch with the most PoW in it, as is already done by Bitcoin~\cite{2015_bitcoin_v0_10_synchronization}.
Third, instead of downloading and processing all historical data, the joining node applies the previously obtained snapshot in good faith to initially fill its UTXO set.
Finally, the joining node fetches and processes the \emph{chaintail}, \ie the remaining full blocks succeeding the snapshot's block height, to finalize synchronizing its UTXO set.
During this full synchronization phase, the joining node additionally inspects the chaintail's coinbase transactions for reaffirmations of its applied snapshot.
If the node learns that its snapshot was the most-reaffirmed one during $\Delta_R$ and was reaffirmed at least $k$ times, it accepts the snapshot, which concludes the bootstrapping step.
Otherwise, the joining node aborts and obtains a different snapshot from another source, \eg by connecting to a new set of neighbors.

\section{Seamless Integration into Bitcoin}%
\label{sec:integration}

\name's main feature is its immediate applicability to Bitcoin~\goalcompatibility.
In this section, we present our means to achieve \emph{gradual opt-in deployability} to Bitcoin via a velvet fork, assuming that a sufficient share of honest miners makes a rational choice to support \name, \eg to preserve storage.

\textbf{On-Chain Data.}
Although snapshot reaffirmations must be publicly announced on Bitcoin's blockchain, \name's utilization of only a block's coinbase field prevents any protocol-breaking changes.
Full nodes that are unaware of \name will ignore any snapshot reaffirmation, and \name nodes will never reject blocks containing incorrect reaffirmations.
Instead, they will try to outpace incorrect reaffirmations with legitimate ones.
Hence, our scheme fulfills the requirements for a gradually deployable velvet fork~\cite{2017_kiayias_velvet_forks,2018_zamyatin_velvet_forks}.
To prevent \name nodes from confusing snapshot reaffirmations with other coinbase data, we propose to encapsulate the reaffirmation accordingly, \eg using a unique prefix and separators such as \code{CoinPrune/[snapshot\_id]/}.

\textbf{Peer-to-Peer Protocol.}
Even though \name allows for external snapshot sources, most joining nodes will likely rely on Bitcoin's network to obtain their initial snapshot.
To enable this Bitcoin-intrinsic snapshot acquisition, we extend Bitcoin's P2P protocol~\cite{2017_bitcoin_p2p} with an additional \code{GETSTATE} message type sent by \name nodes.
Joining nodes send a \code{GETSTATE} message to each neighbor to learn about available recent snapshots.
Each neighbor responds with an inventory (\code{INV} message) that contains the hash values of the snapshot header and the chunks of their most recent available and successfully reaffirmed snapshot as \code{STATE} objects.
The joining node uses these \code{INV} messages to determine which snapshot to obtain and to derive that snapshot's identifier.
Then, the node continues to request individual chunks of the most-advertised snapshot from its neighbors using sequences of \code{GETDATA} messages.
Finally, the node applies the state once all chunks are available.
For increased compatibility, we restrict chunk sizes to \SI{1}{\mega\byte}, \ie Bitcoin's maximum block size, and we introduce a new service flag for Bitcoin's \code{VERSION} handshake to avoid sending unknown messages to \name-unaware nodes.

\textbf{Takeaway.}
Bitcoin nodes can adopt \name immediately without creating forks or causing issues on the P2P layer.

\begin{figure*}[t]
    \begin{minipage}[t]{0.32\textwidth}
        \centering
        \includegraphics{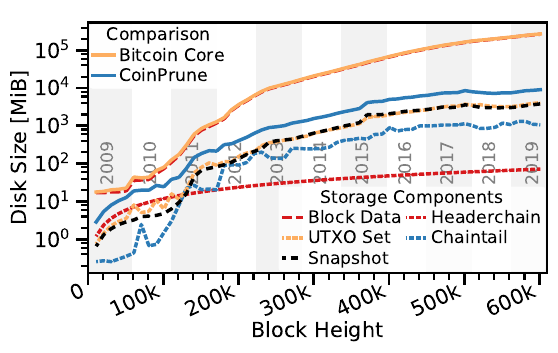}
        \vspace{-.8em}
        \caption{%
            Currently, our scheme already reduces storage requirements by two orders of magnitude.
        }
        \label{fig:eval:disksize}
        \vspace{-1.5em}
    \end{minipage}
    \hfill
    \begin{minipage}[t]{0.32\textwidth}
        \centering
        \includegraphics{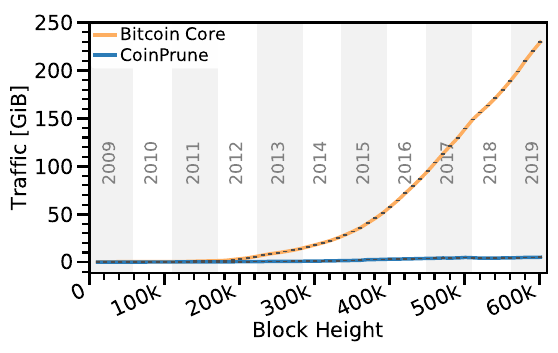}
        \vspace{-.8em}
        \caption{%
            \name allows bootstrapping with vastly reduced amounts of traffic, unburdening all nodes.
        }
        \label{fig:eval:traffic}
        \vspace{-1.5em}
    \end{minipage}
    \hfill
    \begin{minipage}[t]{0.32\textwidth}
        \centering
        \includegraphics{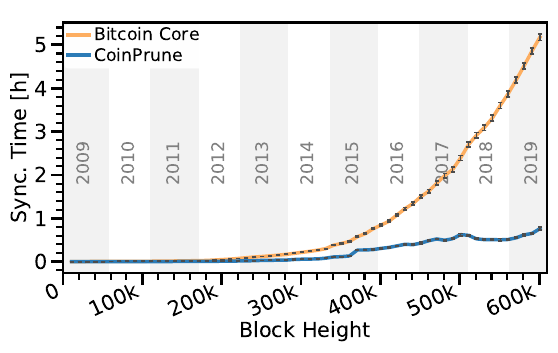}
        \vspace{-.8em}
        \caption{%
			Our scheme reduces initial synchronization times, yet the snapshot size impacts performance.
        }
        \label{fig:eval:sync-time}
        \vspace{-1.5em}
    \end{minipage}
\end{figure*}

\section{Security Discussion}
\label{sec:security}

We argue that \name bootstraps nodes correctly~\goalcorrectness based on verifiable snapshots~\goalverifiability as
\begin{enumerate*}
    \item{our on-chain reaffirmations reliably reference snapshots from arbitrary sources, that}
    \item{positive-only feedback from an honest majority of \name miners establishes trust in snapshots, and that}
    \item{\name is not prone to additional P2P-layer attacks.}
\end{enumerate*}

\textbf{Verifying Snapshot Validity.}
Joining nodes must be able to verify the benignity of a snapshot reliably based on our on-chain reaffirmations.
To achieve this, \name derives snapshot identifiers from a cryptographic hash function in a layered manner.
Each snapshot chunk is hashed individually.
Hence, full nodes cannot alter individual chunks during initial synchronization without having the joining node notice the manipulation.
Furthermore, the snapshot identifier covers the snapshot's meta information as well as all chunks' hash values.
Hence, joining nodes can verify that they obtain exactly the required snapshot chunks, and they know precisely to which block height the snapshot corresponds.
Consequently, no adversary can trick a joining node into accepting an altered snapshot, nor can they create inconsistencies as to how to apply the snapshot (\eg to reintroduce obsolete UTXOs).

\textbf{Reliability of Reaffirmations.}
Even though joining nodes can reliably associate their obtained snapshots with on-chain reaffirmations, our scheme deliberately tolerates the presence of reaffirmations to invalid snapshots to maximize \name's compatibility with Bitcoin.
Consequentially, we need to prevent joining nodes from accepting invalid snapshots as sufficiently reaffirmed.
\name achieves this in a similar manner as Bitcoin keeps its blockchain secure.
Assuming an honest majority among \name-supporting miners, snapshots that are derived correctly from pulse blocks will eventually accumulate reaffirmations faster than other snapshots.
Hence, joining nodes should not rely on snapshots with only fewer than $k$ on-chain reaffirmations.
We note that low adoption of \name could give slow-but-steady malicious miners a relative advantage.
However, by introducing the reaffirmation window $\Delta_R$, adversaries with low relative mining power within the Bitcoin network are highly unlikely to successfully reaffirm an invalid snapshot in time even if they temporarily constitute a dishonest majority among \name nodes.
Finally, a trusted third party can temporarily aid this transition phase by releasing signed snapshot identifiers redundantly to on-chain reaffirmations.

\textbf{P2P Attacks.}
\name integrates well with Bitcoin's P2P protocol and only adds \code{GETSTATE} messages and \code{STATE} inventories.
Hence, considerations regarding Bitcoin's resilience against DoS attacks or Eclipse attacks~\cite{2015_heilman_eclipse_attacks} directly transfer to \name.
Furthermore, adversaries cannot partially manipulate or completely replace valid snapshots, as discussed above.
If joining nodes thus observe an attempted attack, they can abort bootstrapping and connect to a new set of neighbors.

\textbf{Takeaway.}
\name enables joining nodes to obtain snapshots from any source and to still rely on its on-chain reaffirmations to synchronize correctly with the Bitcoin network.

\section{Performance Evaluation}
\label{sec:eval}

We now demonstrate that \name enables massive performance savings for Bitcoin nodes~\goalscalability.
After describing our testbed setup, we present the storage savings achieved for all nodes.
Further, we show that traffic and synchronization time for joining nodes are massively reduced as well.

\subsection{Testbed Setup for Synchronization Measurements}%
\label{sec:eval:setup}

We created a proof-of-concept implementation of \name based on Bitcoin Core v0.17.1.
Our measurements run on a server (2$\times$ Intel Xeon E5-2630 v4, \SI{32}{\giga\byte} RAM, \SI{8}{\tera\byte} Seagate IronWolf ST8000VN0022-2EL112), which synchronizes from eight identical commodity PCs (Intel Core2 Quad Q9400, \SI{8}{\giga\byte} RAM, \SI{500}{\giga\byte} Hitachi Deskstar 7K500) via a Linksys SLM2024 Gigabit switch.
We measure synchronizing with Bitcoin's blockchain in increments of \num{10000} blocks up until a block height of \num{600000} (Oct~19, 2019) and additional \num{1000} blocks (\SI{\sim 1}{week} of blocks) as our chaintail.
We perform synchronization via vanilla Bitcoin Core in one go and start synchronization via \name from the snapshots' respective block heights.
While storage requirements are fully determined by the blockchain data, synchronization times and traffic may vary.
Hence, they were averaged over ten independent runs, and we show the \SI{99}{\percent} confidence intervals.
We omitted to check the coinbase field for the snapshot identifier to be able to use Bitcoin's real blockchain for our measurements.

\subsection{Storage Savings}

\name allows all Bitcoin nodes to prune historical blocks in exchange for maintaining a recent snapshot and the headerchain to serve joining nodes.
In Figure~\ref{fig:eval:disksize}, we depict how the main contributors to Bitcoin's storage demand changed over time in comparison to the serialized snapshot and headerchain required to operate \name.
From this, we derive the overall storage requirements for Bitcoin Core and \name, respectively.
For Bitcoin's storage requirements, we consider the heavily dominating \code{blocks} folder containing raw block data, information required to rewind blocks efficiently, and the block index, as well as the \code{chainstate} folder holding the UTXO set.
In contrast to this, \name needs to store one serialized snapshot and the serialized headerchain, as well as the UTXO set and chaintail for live operation.
Our measurements show that the sizes of serialized snapshots align well with those of Bitcoin's UTXO set.
Minor variances stem from different encodings of both data structures.
Further, persisting the headerchain to reconstruct Bitcoin's block index comes at only negligible costs of \btcstatEvalStorageHeaderchainPerBlock per block, resulting in a headerchain size of \btcstatEvalStorageHeaderchainSize for our latest measurement.
Finally, considering block heights starting from \num{300000}, the chaintail has an average size of \btcstatEvalStorageMeanChaintailSizeLater.
Overall, \name nodes could thus historically reduce their storage requirements by \btcstatEvalStorageMeanReduction, with the largest absolute and relative savings, currently \btcstatEvalStorageMaxSave, at higher block heights.
These savings account for a decrease of two orders of magnitude, with the potential for becoming even larger as the blockchain grows.

\subsection{Evaluation of Synchronization Performance}
\label{sec:eval:performance}

Pruning obsolete data not only relieves Bitcoin nodes from storage depletion but also joining nodes benefit from widespread adoption of \name.
As shown in Figure~\ref{fig:eval:traffic}, the reduced storage requirements directly translate to a reduction in traffic required to synchronize with the Bitcoin network.
For instance, synchronizing from a snapshot on block height \num{600000} with a chaintail length of \num{1000}, joining nodes only inflict \btcstatEvalTrafficBothMaxCompaction of traffic when using \name, whereas legacy nodes would cause \btcstatEvalTrafficBothMaxVanilla of traffic to bootstrap successfully.
Over the whole blockchain, achievable savings average at \btcstatEvalTrafficMeanReduction.
Joining nodes currently have to obtain two orders of magnitude less data during initial synchronization, which is largely dominated by acquiring the snapshot.

A similar trend can be observed for the overall synchronization time of joining nodes, \ie obtaining and verifying the headerchain, the snapshot, and the chaintail.
Figure~\ref{fig:eval:sync-time} shows that \name improves synchronization times over Bitcoin's whole history, resulting in savings of \btcstatEvalSynctimeMeanReduction on average for joining nodes.
Even though Bitcoin mitigates reverifying very old transactions due to its assumed-valid blocks (\cf Section~\ref{sec:sota:updates}), joining nodes still must replay the whole transaction graph.
Contrarily, \name avoids this step as well due to its reliance on snapshots.
In consequence, \name currently enables joining nodes to catch up with the Bitcoin network in \btcstatEvalSynctimeMaxCompaction instead of \btcstatEvalSynctimeMaxVanilla using standard Bitcoin.
This time saving is especially beneficial as initial synchronization is often considered a major scalability concern~\cite{2018_lopp_bitcoin_scalability,2016_croman_blockchain_scalability}.

\textbf{Takeaway.}
The snapshot-based approach of \name unburdens both established and joining nodes from major overhead stemming from Bitcoin's bootstrapping process regarding storage, traffic, and synchronization time.
Hence, \name establishes a secure and effective means to vastly improve Bitcoin's long-term durability.

\section{Conclusion}
\label{sec:conclusion}

\name tackles the increasing scalability issues of public blockchain systems such as Bitcoin.
These issues stem from growing storage and performance issues as nodes keep, redistribute, and reverify obsolete, historical data for security reasons.
We have shown that we can extend Bitcoin with an effective and secure block-pruning scheme without protocol-breaking changes by having honest miners create snapshots of Bitcoin's UTXO set and mutually reaffirm their correctness on the blockchain.
Averaged over Bitcoin's lifetime, joining nodes could have reduced synchronization times by \btcstatEvalSynctimeMeanReductionHighlevel this way.
Furthermore, all nodes are relieved of keeping historical data, reducing Bitcoin's current storage requirements of roughly \btcstatEvalStorageMaxVanillaHighlevel to \btcstatEvalStorageMaxCompactionHighlevel, saving currently \btcstatEvalStorageMaxSaveHighlevel with the prospects of even more saving potential as the blockchain grows.

As future work, we plan to further tackle current limitations of \name and block pruning in general, \eg preserving the currently over \SI{5.3}{\million} application-specific \code{OP\_RETURN} outputs, the efficiency of snapshot creation, and the potential for removing illicit content from the UTXO set.
Hence, \name has the potential to secure Bitcoin's long-term scalability in an immediately applicable, but gradually deployable way.

{
\footnotesize
\vspace{0.5em}
\setstretch{.9}
\textsc{Acknowledgements.}
This work has been funded by the German Federal Ministry of Education and Research (BMBF) under funding reference numbers 16KIS0443, 16DHLQ013, and Z31 BMBF Digital Campus.
The funding under reference number Z31 BMBF Digital Campus has been provided by the German Academic Exchange Service (DAAD).
The responsibility for the content of this publication lies with the authors.\par
}

\vspace{-0.5em}
\bibliographystyle{IEEEtran}
\bibliography{paper}

\end{document}